\documentclass{aa}
\usepackage{graphicx}
\usepackage{psfig}
\usepackage{longtable}
\usepackage{latexsym}
\usepackage{bm}

\begin{document}

\title{Galaxy Evolution in the Environment of ABCG 209}
\titlerunning{Galaxy Evolution in the Environment of ABCG 209}
\author{C. P. Haines\inst{1} 
\and A. Mercurio\inst{2}
\and P. Merluzzi\inst{1}
\and F. La Barbera\inst{1}
\and M. Massarotti\inst{1}
\and G. Busarello\inst{1}
\and M. Girardi\inst{2}
}

\institute{Osservatorio Astronomico di Capodimonte, v. Moiarello 16, I-80131 Napoli, Italy
\and Departimemto di Astronomia, Universit\`{a} degli Studi di Trieste, v. Tiepolo 11, I-34100 Trieste, Italy}
\offprints{C. P. Haines, \email{chris@na.astro.it}}

\date{Received; Accepted}

\def\mpc {\,h^{-1}\,{\rm{Mpc}}}
\def\sqmpc {\,h^{-2}\,{\rm{Mpc}}^{2}}
\def\kpc {\,h^{-1}\,{\rm{kpc}}}
\def\etal  {\it {et al.} \rm}
\def\sqarcmin {\,{\rm arcmin}^{2}}
\def\mgii {\ion{Mg}{ii} }
\def\gsim{ \lower .75ex \hbox{$\sim$} \llap{\raise .27ex \hbox{$>$}}}
\def\lsim{ \lower .75ex \hbox{$\sim$} \llap{\raise .27ex \hbox{$<$}}}
 
\abstract{We examine the environmental effects on the photometric properties of galaxies for the rich galaxy cluster ABCG 209 at $z=0.209$. We use archive CFHT optical imaging of a \mbox{$42\times28\,{\rm arcmin}^{2}$} field centred on the cluster to produce a galaxy sample complete to $B=25.0$ and $R=24.5$. Both the composite and red sequence galaxy luminosity functions are found to be dependent on the local galaxy surface density, their faint-end slopes becoming shallower with increasing density. We explain this as a combination of the morphology-density relation, and dwarf galaxies being cannibalised and/or disrupted by the cD galaxy and the ICM in the cluster core. The $B-R$ colour of the red sequence itself appears 0.02\,mag redder for the highest-density regions, indicative of their stellar populations being marginally ($<5$\%) older or ($<20$\%) more metal-rich. This may be due to the galaxies themselves forming earliest in the rarest overdensities marked by rich clusters, or their star-formation being suppressed earliest by the ICM. 

\keywords{Galaxies: clusters: individual: Abell 209; Galaxies: luminosity function; Galaxies: evolution}
}
\maketitle

\section{INTRODUCTION}

The effect of local environment on the formation and evolution of galaxies is one of the most pressing issues in cosmology today, with evidence that environmental processes affect the mass distribution of galaxies, as well as their star-formation and morphological characteristics. Rich clusters provide a unique opportunity to study these environmental effects, providing large numbers of galaxies at the same redshift which have been exposed to a wide variety of environments. Numerous authors have studied the environmental effects of rich clusters on the luminosity functions, star-formation rates and morphologies of their constituent galaxies (e.g. L\'{o}pez-Cruz \etal \cite{lopezcruz}; Balogh, Navarro \& Morris \cite{balogh}; Hogg \etal \cite{hogg}; Treu \etal \cite{treu}).

The galaxy luminosity function (LF), which describes the number of galaxies per unit volume as a function of luminosity, is a powerful tool for examining galaxy formation and evolution, since it can be directly related to the galaxy mass function. The Press-Schechter prescription for the hierarchical assembly of galaxies predicts a simple analytical formula for the mass distribution of the form \mbox{${\rm n(M)dM}\propto {\rm M}^{\alpha}\exp(-{\rm M/M}^{*})$} (Press \& Schechter \cite{press}) which, despite not considering non-linear dynamical effects, reproduces well the results of N-body simulations of the growth of dark matter halos. This prescription was confirmed by the observation of Schechter (\cite{schechter}) that the galaxy luminosity function can be described by the Schechter function:
\begin{equation}
\psi{\rm (L)\,dL}=\psi^{*}\left(\frac{{\rm L}}{{\rm L}^{*}}\right)^{\alpha}\exp({\rm -L/L}^{*})\,{\rm d}\left(\frac{{\rm L}}{{\rm L}^{*}}\right),
\end{equation}
where the shape of the LF is described by a characteristic cut-off luminosity \mbox{${\rm L}^{*}$}, and $\alpha$, the faint-end slope of the distribution. He also suggested that the shape of the galaxy LF is universal, and only the multiplicative constant $\psi^{*}$ differs between clusters. 

Numerous studies have since been made to determine the galaxy LF: some confirming the validity of the ``{\em universal luminosity function hypothesis}'' (e.g. Lugger \cite{lugger}; Colless \cite{colless}; Trentham \cite{trentham}); whereas others indicate that the LF is instead dependent on the environment of the galaxies, resulting in significant differences between the LFs from cluster to cluster and between cluster and field (e.g. L\'{o}pez-Cruz \etal \cite{lopezcruz}; Valotto \etal \cite{valotto}). More recent studies based on large surveys empirically suggest that the galaxy LF is indeed dependent on environment, with dynamically-evolved, rich clusters and clusters with central dominant galaxies having brighter characteristic luminosities and shallower faint-end slopes, than poorer clusters, those with substructure (L\'{o}pez-Cruz \etal \cite{lopezcruz}; De Propris \etal \cite{depropris}). In particular these differences can be explained through considering the composite LF as the sum of type-specific luminosity functions (TSLFs), each with its universal shape for a specific type of galaxies (e.g. elliptical, spiral, dwarf). The shape of the composite LF then can vary from cluster to cluster, and cluster to field, according to the mixture of different galaxy types in each environment resulting from the morphology-density relation (Dressler \etal \cite{dressler87}; Binggeli \etal \cite{binggeli}).

The galaxies most synonymous with the cluster environment are the bulge-dominated, passively-evolving galaxies which make up the cluster red sequence. Studies of low-redshift clusters  (e.g. Bower, Lucey \& Ellis \cite{bower}) indicate that, irrespective of the richness or morphology of the cluster, all clusters have red sequences, whose k-corrected slopes, scatters and colours are indistinguishable. Other studies find similar precise regularities between other physical properties of the early-type galaxies (surface-brightness, mass-to-light ratio, radius, velocity distribution, metallicity) resulting in the Faber-Jackson (Faber \& Jackson \cite{faber}) and Kormendy (\cite{kormendy}) relations and the fundamental plane (Djorgovski \& Davis \cite{djorgovski}; Dressler \etal \cite{dressler87}). These all indicate that the early-type galaxies which make up the red sequences form a homogeneous population, not only within each cluster, but from cluster to cluster, and also that red sequences are universal and homogeneous features of galaxy clusters, at least at \mbox{$z\lsim0.2$}. 

The photometric evolution of the cluster red sequence with redshift has been studied by a number of authors (e.g. Arag\'{o}n-Salamanca \etal \cite{aragon}; Ellis \etal \cite{ellis}; Stanford \etal \cite{stanford98}; Kodama \etal \cite{kodama98}) for clusters out to \mbox{$z\simeq1.2$} indicating: that red sequence remains a universal feature of clusters; that the stellar populations of its constituent early-type galaxies are formed in a single, short burst at an early epoch \mbox{$(z_{f}\gsim2)$}; and that the galaxy colours have evolved passively ever since. This is confirmed by the spectra of red-sequence galaxies which in nearby clusters are best fit by simple stellar populations of ages \mbox{9--12\,Gyr}, resulting in their having characteristically strong spectral breaks at 4000\,\AA, and correspondingly red \mbox{$U-V$} colours.

Cosmological models of structure formation indicate that the densest regions of the universe corresponding to the rarest overdensities in the primordial density field will have collapsed earliest, and will contain the most massive objects, i.e. rich galaxy clusters. Detailed cosmological simulations which follow the formation and evolution of galaxies (e.g. Kauffmann \cite{kauffmann}; Blanton \etal \cite{blanton00}) predict that galaxies in these high-density regions are older and more luminous than those in typical density (i.e. field) regions. Having initially (\mbox{$z\sim5$}) been the most likely place for the formation of stars and galaxies, the high-density environment of rich cluster cores is later (\mbox{$z\sim$1--2}) filled with shock-heated virialised gas that does not easily cool and collapse (Blanton \etal \cite{blanton99}), inhibiting both the formation of stars and galaxies (Blanton \etal \cite{blanton00}), which instead occurs most efficiently in increasingly less massive dark matter halos. These models are able to successfully predict the observed cluster-centric star-formation and colour gradients, and the morphology-density relation of $z<0.5$ clusters (Balogh \etal \cite{balogh}; Diaferio \etal \cite{diaferio}; Springel \etal \cite{springel}; Okamoto \& Nagashima \cite{okamoto}), although it is not clear whether they predict trends with density of colour or star-formation for galaxies of a fixed morphology and luminosity. They show that these gradients in galaxy properties naturally arise in hierarchical models, because mixing is incomplete during cluster assembly, and the positions of galaxies within the clusters are correlated with the epoch at which they were accreted.

To examine the effect of environment on galaxy properties, in particular the galaxy luminosity function, and the cluster red sequence, we have performed a photometric study of the galaxy cluster A\,209 at \mbox{$z=0.21$} (Kristian, Sandage \& Westphal 1978; Wilkinson \& Oke \cite{wilkinson}; Fetisova \cite{fetisova}; Mercurio \etal \cite{paper1}) using archive wide-field $B$ and $R$-band imaging, which allows the photometric properties of the cluster galaxies to be followed out to radii of \mbox{3--4$h_{70}^{-1}$\,Mpc}. A\,209 is a rich (richness class \mbox{${\rm R}=3$}; Abell \etal \cite{abell}) , X-ray luminous (\mbox{${\rm L}_{{\rm X}}$(0.1--2.4\,keV$)\sim2.7\times10^{45}\,h^{-2}_{70}\,$erg\,s$^{-1}$}, Ebeling \etal \cite{ebeling}; \mbox{${\rm T}_{{\rm X}}\sim10$\,keV}, Rizza \etal \cite{rizza}), and massive cluster (Mercurio \etal \cite{paper1}, but see also Dahle \etal \cite{dahle}). 
The cluster was initially chosen for its richness, allowing its internal velocity field and dynamical properties to be studied in great detail, and also for its known substructure, allowing the effect of cluster dynamics and evolution on the properties of its member galaxies to be examined. 
This substructure is manifested by an elongation and asymmetry in the X-ray emission with two main clumps (Rizza \etal \cite{rizza}), but no strong cooling flow is detected. Moreover, the young dynamical state is indicated by the possible presence of a radio halo (Giovannini, Tordi \& Feretti \cite{giovannini}), which has been suggested to be the result of a recent cluster merger, through the acceleration of relativistic particles by the merger shocks (Feretti \etal \cite{feretti}).

The internal dynamics of the cluster were studied by Mercurio \etal (\cite{paper1}) through a spectroscopic survey of 112 cluster members. A high value of the line-of-sight velocity dispersion was found, with \mbox{$\sigma_{v}=1394^{+88}_{-99}$\,km\,s$^{-1}$}. Assuming dynamic equilibrium, this value of $\sigma_{v}$ leads to a virial radius of \mbox{${\rm R}_{{\rm vir}}\sim2.5h_{70}^{-1}$\,Mpc}, and a virial mass of \mbox{${\rm M(R}<{\rm R}_{{\rm vir}})=2.$3--3.1$\times10^{15}h^{-1}_{70}{\rm M}_{\odot}$}.

Evidence in favour of the cluster undergoing a dynamical evolution is found in the form of a velocity gradient acting along a SE-NW axis, which is the same preferential direction found from the elongation in the spatial distribution of galaxies and X-ray flux, as well as that of the cD galaxy. There is also significant deviation of the velocity distribution from a Gaussian, with evidence for two secondary clumps at \mbox{$z=0.199$} and \mbox{$z=0.215$}, which appear spatially segregated from the main cluster. These all indicate that A\,209 is undergoing strong dynamic evolution with the merging of two or more sub-clumps along the SE-NW direction. This dynamical situation is confirmed by the weak lensing analysis of Dahle \etal (\cite{dahle}) which shows the cluster to be highly elongated in the N-S direction, and two significant peaks, the largest coincident with the cluster centre and cD galaxy, and the second 6\,arcmin to the north.

The cluster was studied previously using optical data covering a \mbox{78\,arcmin$^{2}$} (\mbox{160\,arcmin$^{2}$} in $V$) region, reaching \mbox{$B=22.8$}, \mbox{$V=22.5$} and \mbox{$R=22.0$}, and galaxy luminosity functions determined for each passband (Mercurio \etal \cite{paper2}).

The photometric data are presented in Section 2. As this cluster has been found to be significantly elongated along the SE-NW direction, we examine the effect of the cluster environment by measuring galaxy properties as a function of local surface density rather than cluster-centric radius. We describe the method for determining the cluster environment in Sect. 3. The ability to successfully subtract the background field population is vital to study the global properties of cluster galaxies from photometric data alone, and we describe the approach to this problem in Sect. 4. The composite galaxy LF is presented in Sect. 5, and the properties of the red sequence galaxies described in Sect. 6. The effects of environment on the blue galaxy fraction and mean galaxy colours are presented in Sects. 7 and 8. Sect. 9 is dedicated to the summary and discussion of the results. In this work we assume \mbox{H$_{0}$=70\,km\,s$^{-1}$\,Mpc$^{-1}$}, \mbox{$\Omega_{m}=0.3$}, and \mbox{$\Omega_{\Lambda}=0.7$}. In this cosmology, the lookback time to the cluster A\,209 at \mbox{$z=0.209$} is 2.5\,Gyr.

\section{THE DATA}

The data were obtained from the Canada-France-Hawaii telescope (CFHT) science archive (PI. J.-P. Kneib), comprising wide-field $B$- and $R$-band imaging centred on the cluster A\,209. The observations were made on 14--16 November 1999, using the CFHT12K mosaic camera, an instrument made up of 12 \mbox{$4096\times2048$} CCDs, set at the prime focus of the 3.6-m CFHT. The CCDs have a pixel scale of $0.206^{\prime\prime}$, resulting in a total field of view of \mbox{$42\times28\,{\rm arcmin}^{2}$}, corresponding to \mbox{$8.6\times5.7\,h^{-2}_{70}{\rm Mpc}^{2}$} at the cluster redshift. The total exposure times for both $B$- and $R$-band images are 7\,200s, made up of eight 900s $B$-band and twelve 600s $R$-band exposures, jittered to cover the gaps between the CCDs.

Standard procedures were used to bias-subtract the images, using bias exposures and the overscan regions of each CCD. Saturated pixels and bleed trails from bright stars are identified and interpolated across. The images were then flat-fielded using a superflat made up of science images taken using the same camera/filter setup of cluster A\,209, as well as those of the clusters A\,68, A\,383, A\,963, CL\,0818 and CL\,0819 which were also observed as part of the same observing program. For each image, the sky is subtracted by fitting a quadratic surface to each CCD. The sky is then modelled using a \mbox{$256\times256$} pixel median filter, rejecting all pixels \mbox{$>3\sigma$} from the median (i.e. galaxies). 

Cosmic rays are removed by comparison of each image with a registered reference image (taken to be the previous exposure). After masking off those pixels \mbox{$3\sigma$} above the median in the reference image (taken to be sources) each pixel which has a value \mbox{$3\sigma$} higher than its value in the reference image, is flagged as a cosmic ray and interpolated across. The individual images are corrected for airmass using the prescribed values from the CFHT website of \mbox{$\alpha(B)=-0.17$} and \mbox{$\alpha(R)=-0.06$}. The photometry of non-saturated guide stars is found to agree between exposures to an rms level of 0.003\,mag, indicating the observations were taken in photometric conditions. 

The images are registered and coadded in a two-step process. Each of the 12 CCDs are initially registered and coadded, using the positions of sources from the 2nd Guide Star Catalogue (GSC2) to determine integer-pixel offsets between the jittered exposures. As the exposures are jittered to cover the gaps between the CCDs, there is some overlap between the coadded images of adjacent CCDs. This is used to create linear transforms allowing 12 CCDs to be {\em ``stitched''} together to form an interim coadded image. This interim image is then used as a reference to which each individual exposure is registered. The registered exposures are then coadded after masking out the gaps between CCDs and bad columns, using $3\sigma$ clipping. The distortion produced by the camera optics is modelled and removed as a quartic-polynomial fit to the positions of GSC2 sources in a reference tangential plane astrometry.

The photometric calibration was performed into the Johnson-Kron-Cousins photometric system using observations of $\sim300$ secondary standard stars \mbox{($14<R<17$)} in fields 6 and 7 of Galad\'{\i}-Enr\'{\i}quez, Trullols \& Jordi (\cite{galadi}). The fields also contain equatorial standard stars of Landolt (\cite{landolt}) tying the photometric calibration to the Landolt standards, while allowing the uncertainty of the zeropoint and colour-terms to be reduced and measured in a statistical manner. The median photometric uncertainty for each standard star was \mbox{$\Delta(B)=0.033$}, and \mbox{$\Delta(R)=0.021$}. The zero-points and colour terms were fitted using a weighted least-squares procedure, and are shown in Table~\ref{photometry} along with the observed FWHMs of each image.
\begin{table}
\begin{tabular}{cccc} \hline
    Band FWHM & Colour & Zeropoint & Colour Term \\ \hline
$B$\hspace{0.5cm}1.02 & $B-R$  & $25.720\pm0.008$ & $-0.0180\pm0.005$ \\
$R$\hspace{0.5cm}0.73 & $B-R$  & $25.983\pm0.005$ & $\;\;\,0.0002\pm0.004$ \\ \hline
\end{tabular}
\caption{Photometric parameters of optical data.}
\label{photometry}
\end{table}
Object detection was performed using SExtractor in two-image mode (Bertin \& Arnouts \cite{bertin}) for sources with 4 contiguous pixels 1$\sigma$ over the background level in the $R$-band image. The total $B$ and $R$ magnitudes were taken to be the Kron magnitude, for which we used an adaptive elliptical aperture with equivalent diameter $a.r_{K}$, where $r_{K}$ is the Kron radius, and $a$ is fixed at a constant value of 2.5. $B-R$ colours were determined using fixed apertures of 5\,arcsec diameter (corresponding to \mbox{$\sim17$}\,kpc at \mbox{$z=0.209$}), after correcting for the differing seeing of the $R$- and $B$-band images. The measured magnitudes were corrected for galactic extinction following Schlegel, Finkbeiner \& Davis (\cite{schlegel}) measured as \mbox{${\rm E}(B-V)=0.019$}, giving \mbox{${\rm A}(B)=0.083$} and \mbox{${\rm A}(R)=0.051$}. The uncertainties in the magnitudes were obtained by adding in quadrature both the uncertainties estimated by SExtractor and the uncertainties of the photometric calibrations. 

\begin{figure}[t]
\centerline{{\resizebox{\hsize}{!}{\includegraphics{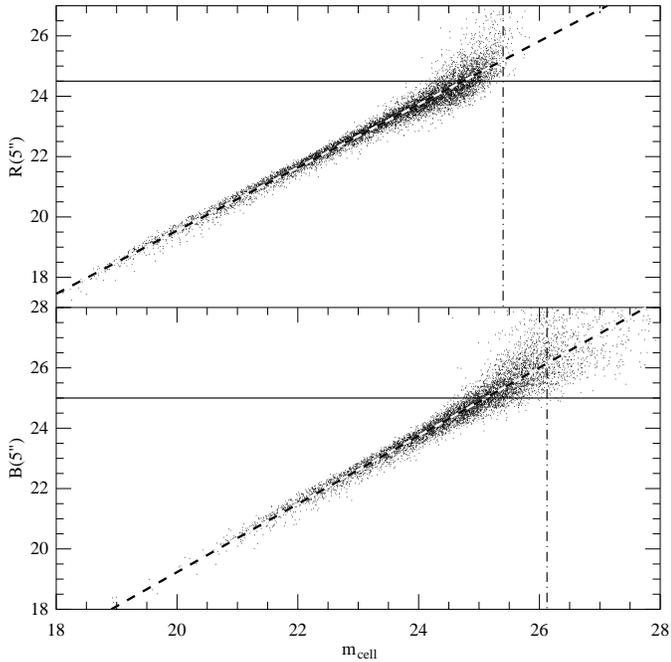}}}}
\caption{Completeness magnitudes of the $B$- and $R$-band images, as estimated by comparing the magnitudes in the fixed aperture (m$_{ap}$) and in the detection cell (m$_{cell}$). The horizontal solid lines represent the completeness limits, the vertical dot-dashed lines mark the limits in the detection cell, and the dashed lines are the linear relations between m$_{ap}$ and m$_{cell}$.}
\label{maglimit}
\end{figure}

The completeness magnitudes were derived following the method of Garilli, Maccagni \& Andreon (\cite{garilli}), as shown in Fig.~\ref{maglimit}, which compares the magnitudes in the fixed aperture (m$_{ap}$) and in the detection cell (m$_{cell}$) which in our case is an aperture of area 4\,pixels. The magnitude limits at which galaxies are lost due to being fainter than the threshold in the detection cell are determined, and are indicated by the vertical dot-dashed lines.
There is a correspondence between $m_{cell}$ and m$_{ap}$ as shown by the dashed line, which has a certain scatter that depends essentially on the galaxy profile and the photometric errors. To minimise biases due to low-surface brightness galaxies, the completeness magnitude limits are chosen to consider this dispersion, and are indicated by the horizontal solid lines at \mbox{$B=25.0$} and \mbox{$R=24.5$}. For the analyses of the cluster red sequence, we consider a magnitude limit of \mbox{$R=23.0$}, in order that we remain complete for galaxies with \mbox{$B-R\sim2.0$}.

\subsection{Star - Galaxy Separation}

\begin{figure}[t]
\centerline{{\resizebox{6cm}{!}{\includegraphics{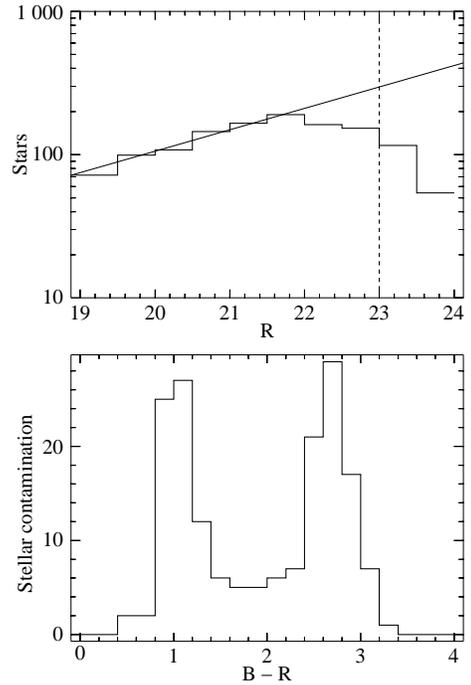}}}}
\caption{Photometric properties of sources classified by SExtractor as being stars. {\bf (top panel)} Number-magnitude distribution of stars. The best-fitting power-law function to the counts for \mbox{$19<R<22$} is indicated, as is the $R=23$ magnitude limit used for analysis of the red sequence galaxy population. {\bf (bottom panel)} $B-R$ colour distribution of \mbox{$19<R<22$} stars, normalised to match the expected level of stellar contamination for \mbox{$22<R<23$}.} 
\label{stars}
\end{figure}

Star-galaxy separation is performed using the SExtractor stellarity index, with stars defined as sources with stellarities \mbox{$\geq0.98$}. By examination of the distributions of sources classified as stars and galaxies in the magnitude-FWHM plane, and the number-magnitude distribution of sources classified as stars, we believe the classification to be efficient to \mbox{$R\sim22$} and \mbox{$B\sim23$}, where stars constitute 12\% of all sources. 

\begin{figure*}[t]
\centerline{{\resizebox{\hsize}{!}{\includegraphics[angle=-90]{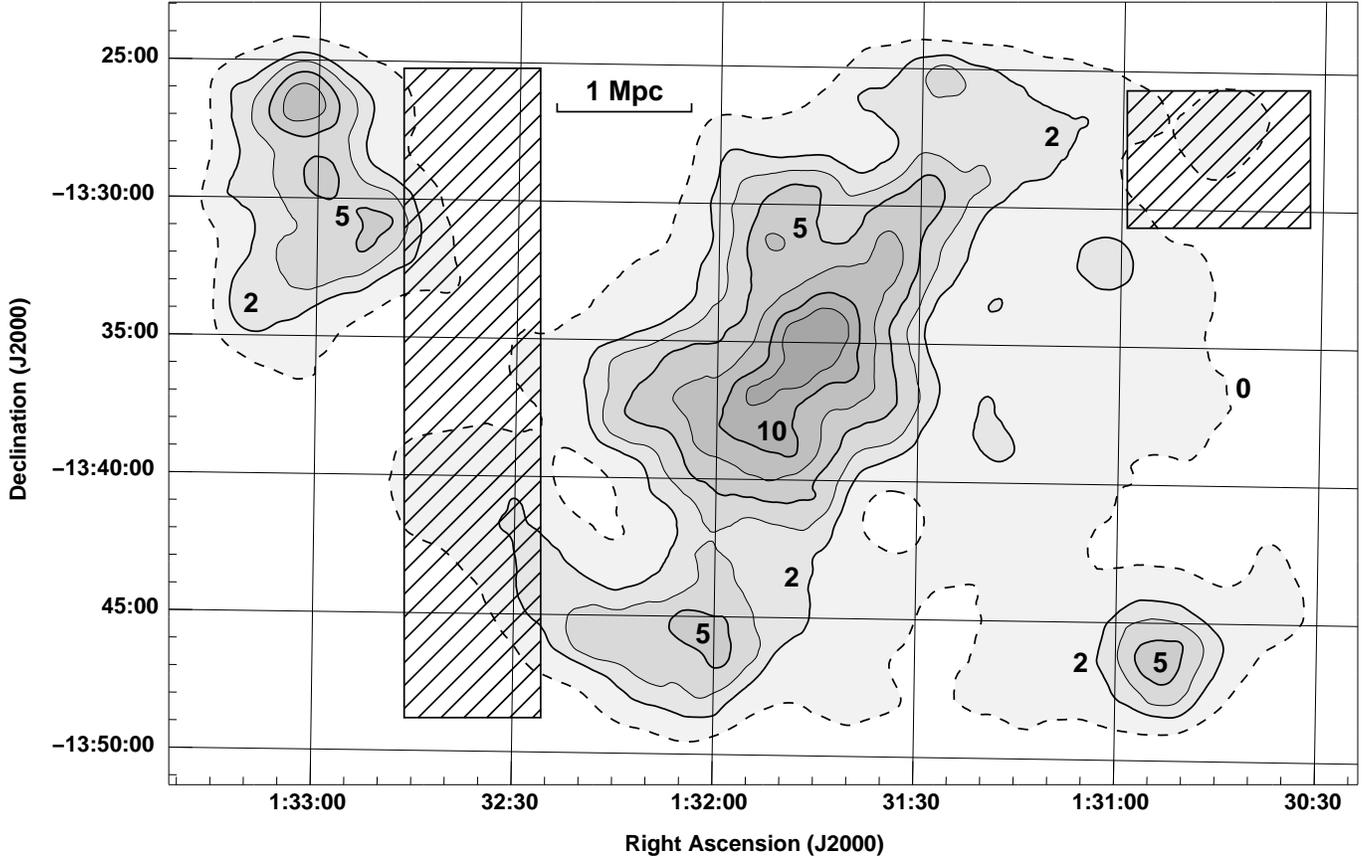}}}}
\caption{The surface density of $R<23.0$ galaxies in the field of A\,209. The dashed contour corresponds to the field galaxy number density, while the solid contours correspond respectively to 2, 3.3, 5, 7.5, 10, and 12.5 cluster galaxies\,arcmin$^{-2}$. The shaded boxes indicate the regions used to represent the field galaxy population.}
\label{density}
\end{figure*}

Figure~\ref{stars}(top) shows the number-magnitude distribution of sources classified as stars by SExtractor. For \mbox{$19<R<22$} the distribution behaves as a power-law as would be expected (see Groenewegen \etal \cite{groenewegen}). At brighter magnitudes, stars become saturated, while at fainter magnitudes, the classifier begins to misclassify stellar sources as galaxies. By extrapolating the power-law to fainter magnitudes, as indicated by the solid diagonal line, we estimate the level of contamination by stars in the range \mbox{$22<R<23$} to be \mbox{$\approx200$} over the whole field (0.18\,stars\,arcmin$^{-2}$), or $\sim5\%$ of all sources. However, since the contamination from faint stars should affect those regions used to define the field galaxy population at the same level as for the cluster regions, when we correct for field galaxy contamination, the contamination due to faint stars should automatically cancel out also. 

Figure~\ref{stars}(bottom) shows the $B-R$ colour distribution of stars in the range \mbox{$19<R<22$}. The y-axis is normalised to represent the expected level of contamination by stars in the range \mbox{$22<R<23$} over the whole field, i.e. 200 in total, assuming the colour distribution remains constant. The distribution is clearly bimodal with two sharp peaks at \mbox{$B-R\sim1.0$} and \mbox{$B-R\sim2.7$} indicating the dominant populations of white and red dwarf stars (Groenewegen \etal \cite{groenewegen}). Fortunately, as regards the study of the red sequence galaxy population of A\,209, at these magnitudes the red sequence corresponds approximately to \mbox{$2.0<B-R<2.2$}, midway between the two peaks in the colour distribution of stars. Hence, the expected level of contamination by stars to the red sequence galaxy population is minimal, corresponding to \mbox{$\sim6$} stars over the whole mosaic image with \mbox{$22<R<23$}.

\section{Defining the Cluster Environment}

To study the effect of the cluster environment on galaxies in the vicinity of A\,209, the local surface density of galaxies, $\Sigma$, is determined across the CFHT images. This is achieved using an adaptive kernel estimator (Pisani \cite{pisani93}; \cite{pisani96}), in which each galaxy is represented by an exponential kernel, \mbox{$K(r)\propto\exp(-r/r_{0})$}, whose width, $r_{0}$ is proportional to $\Sigma^{-1/2}$, thus ensuring greater resolution where it is needed in the high-density regions, and more smoothing in the low-density regions where the signal-to-noise levels are much lower. For this study, the surface number density of \mbox{$R<23.0$} galaxies is considered, the magnitude limit to which cluster red sequence galaxies can be detected in both passbands. The local density for each galaxy is initially determined using an exponential kernel whose width is fixed to 60\,arcsec, and then iteratively recalculated using adaptive kernels, before corrected for field contamination. The resultant surface number density map of the A\,209 field is shown in Fig.~\ref{density}. The dashed contour corresponds to the density of field galaxies, while the thick contours correspond to 2, 5 and 10\, cluster galaxies arcmin$^{-2}$, the densities used to separate the three cluster environments described below. 

The spatial distribution shows the complex structure of this cluster, characterised by the clear elongation in the SE-NW direction (see Mercurio \etal \cite{paper1}), indicative of the cluster having gone through a recent merger event. There are also present three other subclumps, the first one about 10\,arcmin south along the cluster elongation, the second at about 18\,arcmin to the north-east, and the last one, the smallest, at about 18\,arcmin to the south-west. The colour information on galaxies in these peaks confirm that these could be groups/clusters at the same redshift as A\,209, each peak containing galaxies belonging to the cluster red sequence of A\,209. Given the complexity of the structure of A\,209, it is important that to study the effect of environment on galaxy properties, we measure them as a function of local galaxy number surface density rather than cluster-centric radius as is usual in these type of studies.

For the following analyses on the effect of the cluster environment on its constituent galaxy population we define three regions selected according to the local surface number density. Firstly we consider a high-density region with \mbox{$\Sigma>10\,$gals arcmin$^{-2}$}, which corresponds to the cluster core out to a radius of \mbox{$\sim$500\,kpc}, and covers \mbox{16.0\,arcmin$^{2}$}. Next we consider intermediate- \mbox{($5<\Sigma<10\,$gals arcmin$^{-2}$)} and low-density \mbox{($2<\Sigma<5\,$gals arcmin$^{-2}$)} regions which probe the cluster periphery with median galaxy cluster-centric radii of 1.3 and 2.2\,Mpc, and cover 68.0 and \mbox{189.6\,arcmin$^{2}$} respectively. It should be stated that in each case we are studying the cluster galaxy population and not that of the field, and the low-density environment represents an overdense region.

\section{Statistical Field Galaxy Subtraction}

\begin{figure}[t]
\centerline{{\resizebox{\hsize}{!}{\includegraphics{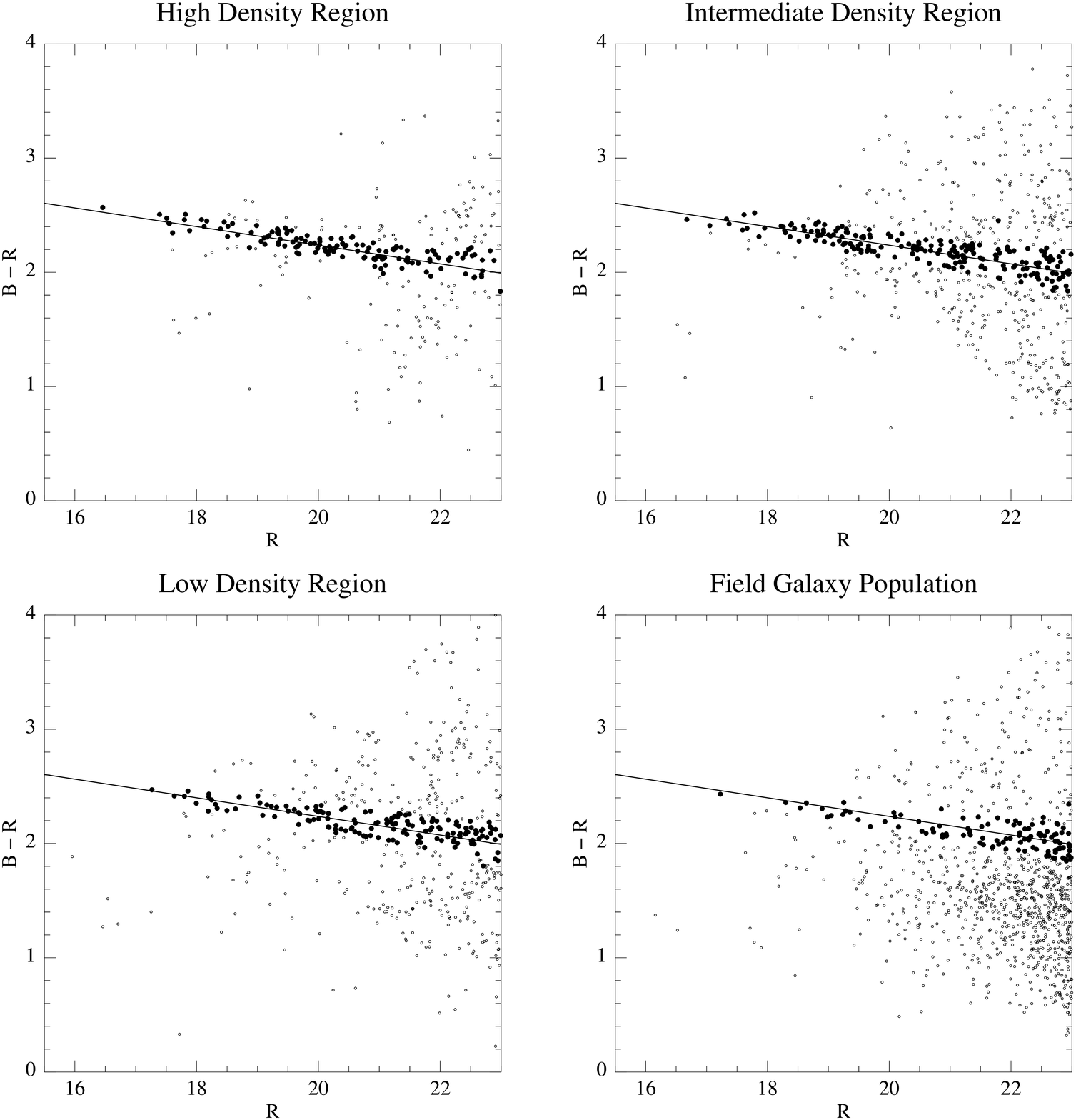}}}}
\caption{The \mbox{$B-R/R$} colour-magnitude diagrams of the cluster galaxy population in the three cluster regions corresponding to high-, intermediate-, and low-density environments. Each of the three diagrams is the result of a Monte-Carlo realisation in which field galaxies are subtracted statistically. For comparison the C-M diagram of field galaxies is shown also. In each plot the solid line indicates the best-fitting C-M relation of Eq.~\ref{CMrelation}, and galaxies identified as belonging to the red sequence through Eq.~\ref{caustic} are indicated by solid circles.}
\label{CMs}
\end{figure}

To accurately measure the properties of the cluster population as a function of environment requires the foreground / background contamination to be estimated efficiently and corrected. Given the lack of spectroscopic information, we estimate the field galaxy contamination as a function of magnitude and colour (see also Kodama \& Bower \cite{kodama}).

The sample of field galaxies is taken from two regions within the CFHT12K field indicated in Fig.~\ref{density} by the shaded boxes. These areas were identified as regions where the local galaxy surface density reached a stable minimum level over a wide area indicative of the field.  They also have galaxy number counts within $1\sigma$ of the ESO-Sculptor survey of Arnouts et al. (\cite{arnouts}) which covers an area of \mbox{$0.24\times1.52\,$deg$^{2}$}, and is complete to \mbox{$B=24.5$} and \mbox{$R=23.5$}. The two {\em ``field''} regions in the A\,209 field cover in total \mbox{160.9\,arcmin$^{2}$}, and contain 965 galaxies to \mbox{$R=23.0$}, resulting in a number surface density of \mbox{6.0\,gals.arcmin$^{-2}$}, as indicated in Fig.~\ref{density} by the dashed contour. 

A two-dimensional distribution histogram of these field galaxies is built with bins of width 0.25\,mag in \mbox{$B-R$} colour and 0.5\,mag in $R$ magnitude. Similar histograms are the constructed for each of the three cluster regions, which also obviously include a field galaxy contamination. The field histogram is normalised to match the area within each cluster region, and for each bin \mbox{$(i,j)$} the number of field galaxies \mbox{$N_{i,j}^{field}$} and the cluster plus field \mbox{$N_{i,j}^{cluster+field}$} are counted. The former can be larger than the latter for low-density bins due to low number statistics, in which case the excess number of field galaxies are redistributed to neighbouring bins with equal weight until \mbox{$N_{i,j}^{field}<N_{i,j}^{cluster+field}$} is satisfied for all bins. For each galaxy in the cluster region the probability that it belongs to the field is then defined as:
\begin{equation}
P(field)=\frac{N_{i,j}^{field}}{N_{i,j}^{cluster+field}},
\end{equation}
where the numbers are taken from the bin that the galaxy belongs to. For each cluster region, 100 Monte-Carlo simulations of the cluster population are realised and averaged. 

The \mbox{$B-R/R$} C-M diagrams for Monte-Carlo realisations of the cluster galaxy population for each of the three cluster regions corresponding to high-, intermediate-, and low-density environments are shown in Fig.~\ref{CMs}, along with that for the field region used for the statistical field subtraction. The different distributions of the cluster and field populations are obvious, with the red sequence prominent in both high- and intermediate-density regions.
 
\section{The Galaxy Luminosity Function}

\begin{figure}[t]
\centerline{{\resizebox{\hsize}{!}{\includegraphics{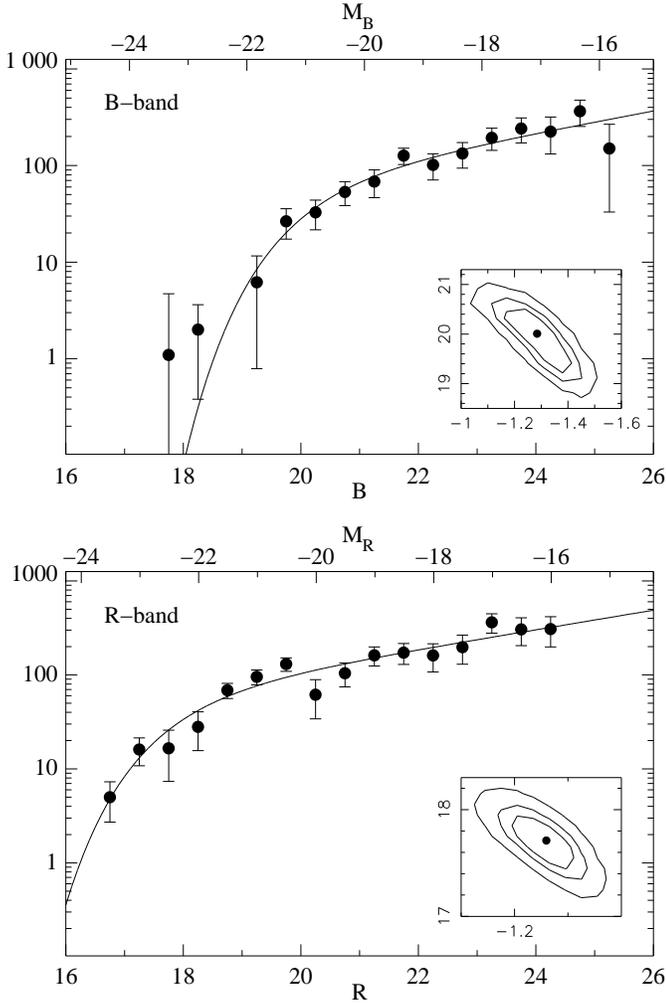}}}}
\caption{The $B$- and $R$-band LFs of galaxies in the virial region. The solid curve indicates the best-fitting Schechter function whose parameters are indicated in Table~\ref{schechter}. In the small panels, the 1, 2 and $3\sigma$ confidence levels of the best-fitting parameters $\alpha$ and $M^{*}$ are shown.}   
\label{lumfuncs}
\end{figure}

\begin{figure}[t]
\centerline{{\resizebox{\hsize}{!}{\includegraphics[angle=-90]{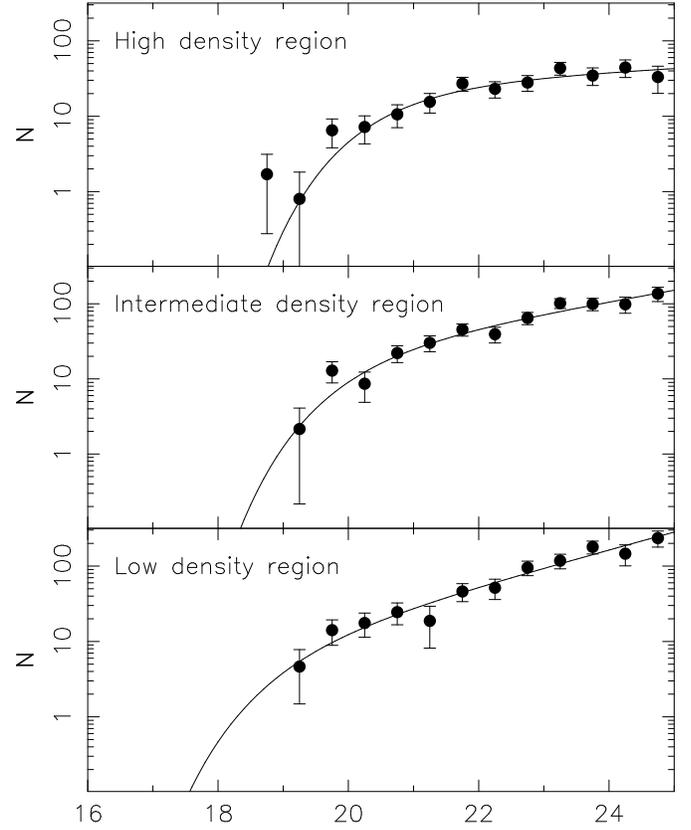}}}}
\caption{The $B$-band LFs of galaxies in the three cluster regions corresponding to high-, intermediate- and low-density environments.}
\label{Blumfuncs}
\end{figure}

\begin{figure}
\centerline{{\resizebox{7cm}{!}{\includegraphics{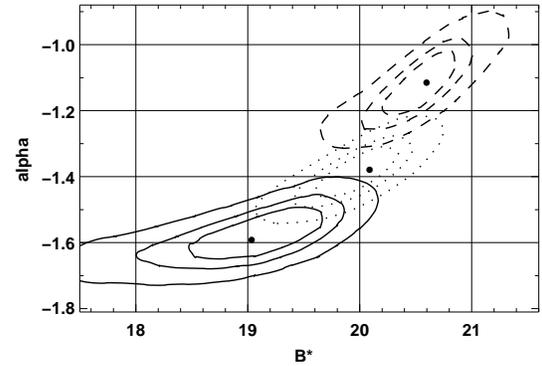}}}}
\caption{The 1, 2 and $3\sigma$ confidence limits for the best-fitting Schechter parameters $\alpha$ and $B^{*}$ for the three cluster regions corresponding to low- (solid contours), intermediate- (dotted) and high-density (dashed) environments.}
\label{contB}
\end{figure}

\begin{figure}[t]
\centerline{{\resizebox{\hsize}{!}{\includegraphics[angle=-90]{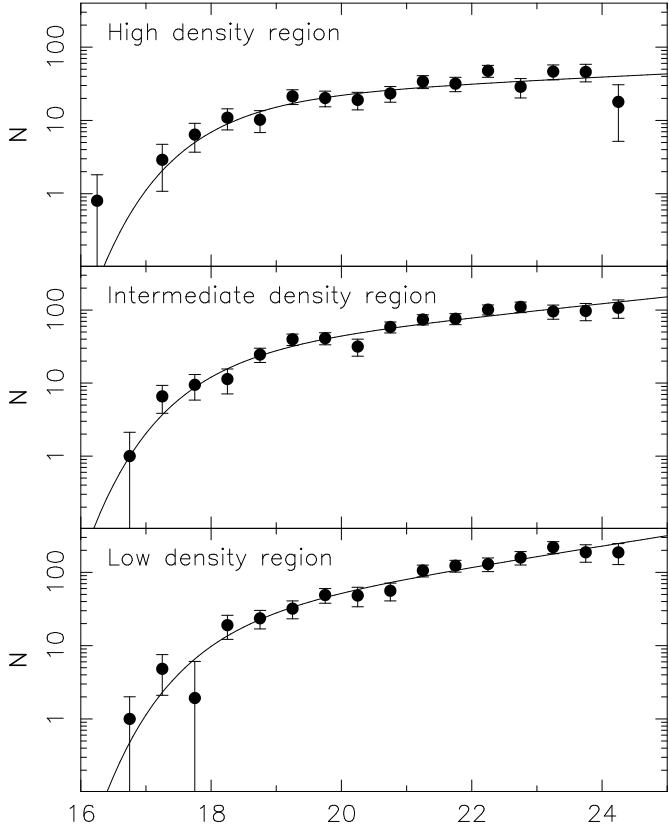}}}}
\caption{The $R$-band LFs of galaxies in the three cluster regions corresponding to high-, intermediate- and low-density environments.}
\label{Rlumfuncs}
\end{figure}

\begin{figure}
\centerline{{\resizebox{7cm}{!}{\includegraphics{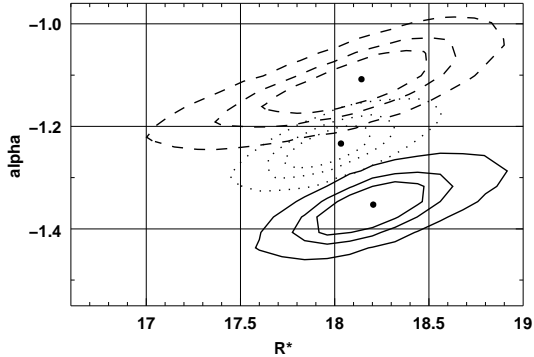}}}}
\caption{Confidence limits for the best-fitting Schechter parameters $\alpha$ and $R^{*}$ for the three cluster regions. Contours as for Fig.~\ref{contB}.}
\label{contR}
\end{figure}

In order to measure the cluster LF in each band we used all the galaxy photometric data up to the completeness magnitudes and removed the interlopers by statistically subtracting the background contamination, as determined from the two field regions. The errors on the cluster LFs are computed by adding in quadrature the Poisson fluctuations for the galaxy counts in both the cluster and field regions, and fluctuations due to photometric errors on Kron magnitudes.

We compute the cluster LFs for both $B$ and $R$ bands for galaxies within one virial radius, and examine the effect of the cluster environment by determining the LFs for each of the three cluster regions corresponding to \mbox{high-,} intermediate-  and low-density environments. For each cluster LF, we fit the observed galaxy counts with a single Schechter function. Absolute magnitudes are determined using the k-corrections for early-type galaxies from Poggianti (\cite{poggianti}). All the fit parameters and associated $\chi^{2}$ statistics are listed in Table~\ref{schechter}.

\begin{table}
\begin{center}
\begin{tabular}{cccccc} \hline
 & Region & $B^{*}$ & M$^{*}$ & $\alpha$ & $\chi^{2}_{\nu}$ \\ \hline 
$\!\!B\!\!$ & $r<{\rm R}_{vir}$         & 20.01 & -21.07 & -1.28 & 0.59 \\ \hline
$\!\!B\!\!$ & $\Sigma>10$               & 20.60 & -20.48 & -1.11 & 0.61 \\
$\!\!B\!\!$ & $\!5\!<\!\Sigma\!<\!10\!$ & 20.09 & -20.99 & -1.38 & 0.92 \\
$\!\!B\!\!$ & $\!2\!<\!\Sigma\!<\!5\!$  & 19.03 & -22.05 & -1.50 & 0.65 \\ \hline

 & Region & $R^{*}$ & & &  \\ \hline 
$\!\!R\!\!$ & $r<{\rm R}_{vir}$         & 17.71 & -22.55 & -1.26 & 1.07 \\ \hline
$\!\!R\!\!$ & $\Sigma>10$               & 18.14 & -22.12 & -1.11 & 0.85 \\
$\!\!R\!\!$ & $\!5\!<\!\Sigma\!<\!10\!$ & 18.03 & -22.23 & -1.23 & 0.88 \\
$\!\!R\!\!$ & $\!2\!<\!\Sigma\!<\!5\!$  & 18.20 & -22.06 & -1.35 & 0.66 \\ \hline

\end{tabular}
\end{center}
\caption{Fits to the LFs for cluster galaxies. Errors on the $M^{*}$ and $\alpha$ parameters are indicated by the confidence contours shown in Figures~\ref{lumfuncs},~\ref{contB} and~\ref{contR}.}
\label{schechter}
\end{table}

\subsection{Galaxies within the Virial Radius}

Figure~\ref{lumfuncs} shows the LFs in the $B$ and $R$ bands for
galaxies within one virial radius (\mbox{$2.5h^{-1}_{70}$\,Mpc}) of the cluster
centre. The solid curves show the best-fitting single Schechter
functions, obtained by weighted parametric fits to the statistically
background-subtracted galaxy counts. According to the $\chi^{2}$ statistic, the
global distributions of the data are well described by single Schechter functions, although there is an indication of a dip at \mbox{$R\sim20.5$}, as already noted in the previous study of the LF in the central region of the cluster (Mercurio \etal \cite{paper2}).

In that study, where a smaller area was considered, the amplitude of this dip was calculated by comparing expected and observed counts in the range \mbox{$R=2$0--21}, and found to be \mbox{${\rm A}=16\pm8$\%}. For galaxies within the virialised region we obtain a value for the dip amplitude of \mbox{${\rm A}=32\pm20$\%}, which is consistent with that previously obtained.

In Mercurio \etal (\cite{paper2}) we derived the LFs for the central field of \mbox{$9.2^\prime \times 8.6^\prime$} (\mbox{$1.9\times1.8h^{-2}_{70}\,{\rm Mpc}^{2}$}) by using EMMI--NTT images, complete to \mbox{$B=22.8$} and \mbox{$R=22.0$}. In this previous study we subtracted statistically the field contamination using background counts in $B$- and $R$-bands from the ESO-Sculptor Survey (Arnouts \etal \cite{arnouts}; de Lapparent \etal \cite{delapparent}). The parameters of the best-fitting Schechter functions were: \mbox{$B^{*}=20.06$}, \mbox{$\alpha_{B}=-1.26$}, and \mbox{$R^{*}=17.78$}, \mbox{$\alpha_{R}=-1.20$}, which are fully consistent with those derived here.

\subsection{The effect of environment}

To examine the effect of environment on the galaxy LF, we have determined the $B$- and $R$-band LFs for galaxies in three regions selected according to their local density. Figures~\ref{Blumfuncs} and~\ref{Rlumfuncs} show respectively the $B$- and $R$-band LFs in the three different cluster regions, corresponding to high-, intermediate-, and low-density environments. Each LF is modelled through use of a weighted parametric fit to a single Schechter function, the results of which are presented in Table~\ref{schechter}. %The effect of environment on $M^{*}$ is measured by fixing $\alpha$ to the value determined for the region within one virial radius, and fitting a Schechter function to the bright end of the galaxy LF ($B<22$ and $R<20$) for each region. 

As for the region within one virial radius, the single Schechter function gives a good representation of the global distribution of the data for each cluster environment in both $B$- and $R$-bands. In both high- and low-density regions, no dip is apparent at \mbox{$R=2$0--21}, although the observed counts are marginally below that predicted from the Schechter function. Only in the intermediate-density region is a significant dip apparent, where a deficit of \mbox{$R=20.$0--20.5} galaxies significant at the $2\sigma$ level is observed with respect to the fitted Schechter function.

Figures~\ref{contB} and~\ref{contR} show the confidence contours for the best fitting Schechter functions for $\alpha$ and $B^{*}$, and $\alpha$ and $R^{*}$ respectively, for each of the three cluster regions, allowing the trends with density to be followed. In both $B$ and $R$ bands, the faint-end slope becomes significantly steeper from high- to low-density environments, the values of $\alpha$ determined for the high- and low-density regions being inconsistent at more than the $3\sigma$ confidence level.

\section{Properties of the Red Sequence Galaxies}

We determined the colour-magnitude (CM) relation of galaxies in the cluster A\,209 by averaging over 100 Monte-Carlo realisations of the cluster population, and fitting the resultant photometric data of the $\sim480$ $R<21$ galaxies within one virial radius with the biweight algorithm of Beers \etal (\cite{beers}) obtaining
\begin{equation}
(B-R)_{CM} = 3.867\pm0.006 - 0.0815\pm0.0098\times R.
\label{CMrelation}
\end{equation}
Here the errors quoted consider the uncertainty due to background subtraction, photometric errors, and uncertainty from the fitting itself as measured using the scale value of the biweight algorithm. The $B-R$ colour dispersion around the red sequence is measured as a function of magnitude, $\sigma(R)$, for galaxies in each magnitude bin over the range $18<R<23$. It is found to be consistent with being due to an intrinsic colour dispersion, $\sigma_{int}$ and the photometric error, $\sigma_{B-R}^{2}(R)$ added in quadrature,
\begin{equation}
\sigma(R)^{2}=\sigma_{int}^{2}+\sigma_{B-R}^{2}(R).
\end{equation}
The intrinsic dispersion around the red sequence is found to be $\sigma_{int}=0.10$\,mag, which is much higher than the dispersion level of 0.05\,mag found for the Coma and Virgo clusters (Bower, Lucey \& Ellis \cite{bower}), and which is often quoted as being typical for low redshift rich clusters. However, in a similar study of the $B-R/R$ CM-relation for 11 $0.07<z<0.16$ clusters Pimbblet \etal (\cite{pimbblet}) obtain typical dispersion levels of 0.06--0.08, and observe three clusters with dispersion levels higher than ours.

Using this equation, we then defined those galaxies lying within the region between the curves:
\begin{equation}
(B-R)_{\pm} = (B-R)_{CM} \pm \sqrt{\sigma^{2}_{int} + \sigma^{2}_{B-R}}
\label{caustic}
\end{equation}
as being red sequence galaxies. %Note that the criteria for selecting red sequence galxies differs from that used previously in Mercurio \etal (\cite{paper2}) in order to account for the large dispersion level observed. It should be stated though, that this change does not affect the results or trends described below, except to increase the sample size, and hence reduce the size of the error bars.

\begin{figure}[t]
\centerline{{\resizebox{\hsize}{!}{\includegraphics{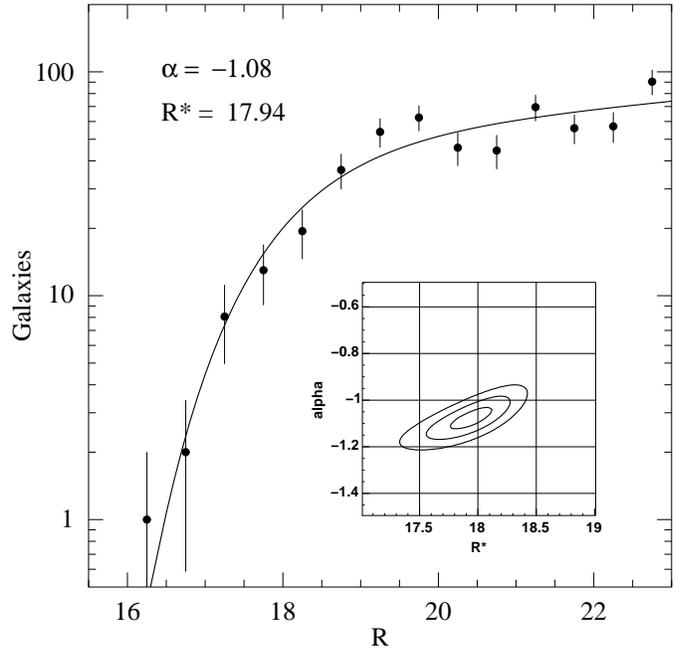}}}}
\caption{The $R$-band LF of red sequence galaxies in the virial region. The solid curve indicates the best-fitting Schechter function whose parameters are indicated in the top-left of each plot. In the small panels, the 1, 2 and $3\sigma$ confidence levels of the best-fitting parameters $\alpha$ and $R^{*}$ are shown.}   
\label{CMlum}
\end{figure}

As for the composite luminosity function, we determine the $R$-band luminosity function of red sequence galaxies within one virial radius of the cluster centre, as shown in Figure~\ref{CMlum}.

\begin{figure*}[t]
\centerline{{\resizebox{\hsize}{!}{\includegraphics{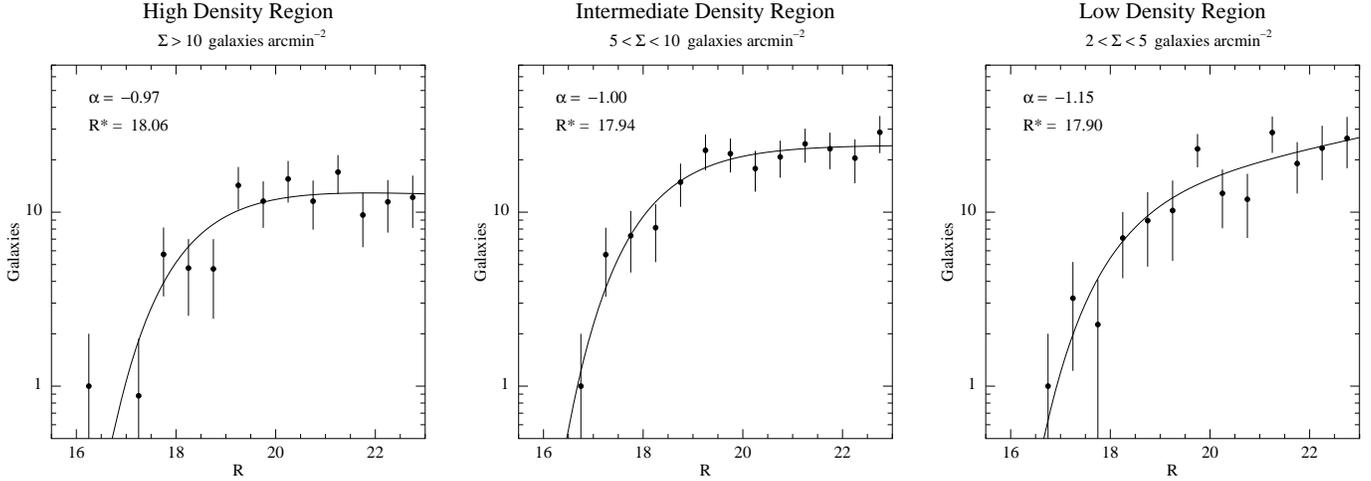}}}}
\caption{The $R$-band LFs of red sequence galaxies in the three cluster regions corresponding to high-, intermediate- and low-density environments. The solid curve indicates the best-fitting Schechter function whose parameters are indicated in the top-left of each plot.}
\label{CMlums}
\end{figure*}

\begin{figure}[t]
\centerline{{\resizebox{6cm}{!}{\includegraphics{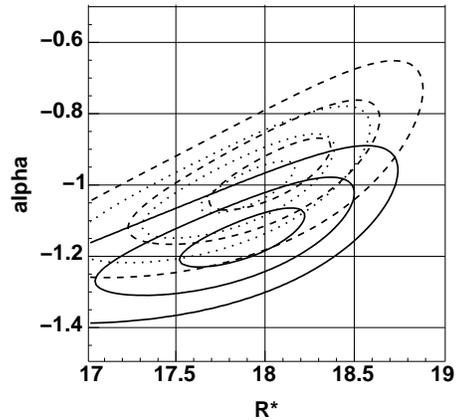}}}}
\caption{Confidence limits for the best-fitting Schechter parameters $\alpha$ and $R^{*}$ for the three cluster regions. Contours as for Fig.~\ref{contB}.}
\label{max}
\end{figure}

\subsection{The effect of environment on the C-M relation}

We also determine the C-M relation in the high- and intermediate-density regions separately, to see if there is any environmental effect on the relation. By fixing the slope at -0.0815, we find the red sequence to be $0.022\pm0.014$\,mag redder in the high-density region than for the intermediate-density region. The effect of environment on the red sequence slope was also examined by leaving the slope as a free parameter, but no significant change was observed. It was not possible to extend this analysis to the low-density region, as the red sequence was poorly constrained due to a much larger contamination by outliers. 

The $R$-band luminosity functions of red sequence galaxies (as selected according to Eq.~\ref{caustic}) in each of the three cluster environments is shown in Figure~\ref{CMlums}, and the best fitting parameters shown in Table~\ref{CMlumparams}. The solid curve shows the best-fitting Schechter function found through a maximum likelihood analysis whose parameters $\alpha$ and $R^{*}$ are indicated in the top-left of each plot. The errors for each bin are determined as the uncertainty from the field galaxy subtraction and the Poisson noise due to galaxy counts in the cluster and field regions, all added in quadrature. 

\begin{table}
\begin{center}
\begin{tabular}{ccccc}\hline
Region & $R^{*}$ & $M^{*}$ & $\alpha$ & $\chi_{\nu}^{2}$\\ \hline
$r<{\rm R}_{vir}$ & 17.94 & -22.22 & -1.08 & 1.61\\ \hline
$\Sigma>10$   & 18.04 & -22.30 & -0.97 & 0.95\\
$5<\Sigma<10$ & 17.94 & -22.20 & -1.00 & 0.45\\
$2<\Sigma<5$  & 17.90 & -22.24 & -1.15 & 0.90\\ \hline
\end{tabular}
\end{center}
\caption{Fits to the red sequence galaxy LFs. Errors on the $R^{*}$ and $\alpha$ parameters are indicated by the confidence contours shown in Figs.~\ref{CMlum} and~\ref{max}.}
\label{CMlumparams}
\end{table}

For the cluster red sequence we consider a magnitude limit of $R=23.0$. At this magnitude, red sequence galaxies have $B-R\sim2.0$, and so are at the completeness limit of the $B$-band image, and have typical uncertainties of $\Delta(R)\sim0.06$ and $\Delta(B-R)\sim0.11$. As discussed earlier, the expected contamination of stars is minimal due to their colour-distribution, and for the three cluster regions corresponding to high-, intermediate-, and low-density environments, should be at the level of 0.1, 0.3 and 0.8 stars respectively. 

Figure~\ref{max} shows the confidence contours for the best-fitting Schechter parameters $\alpha$ and $R^{*}$ for each of the three cluster regions, allowing the trends with density to be followed. As for the overall galaxy luminosity function, the $R$-band luminosity function of red sequence galaxies has a steeper faint-end slope, $\alpha$, in the low-density regions than for their high-density counterparts, although the effect here is smaller, and significant only at the $1.7\sigma$ level. Also in each of the three cluster environments, the faint-end slope for the overall galaxy luminosity function is steeper than that for the luminosity function of red sequence galaxies only. This is due to the red sequence galaxies dominating at bright magnitudes, while at faint magnitudes they constitute only $\sim25\%$ of the cluster population.

\section{Blue Galaxy Fraction}

\begin{figure}[t]
\centerline{{\resizebox{\hsize}{!}{\includegraphics{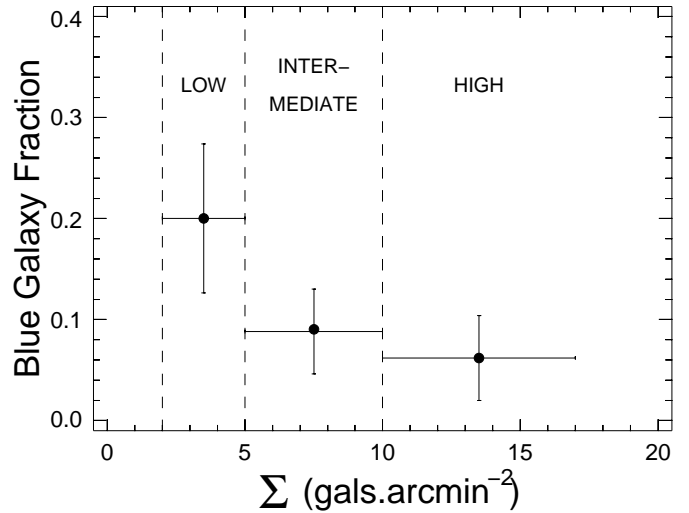}}}}
\caption{The blue galaxy fraction as a function of local number density.}
\label{blue}
\end{figure}

One of the classic observations of the study of galaxy evolution is the Butcher-Oemler (\cite{bo}) effect in which the fraction of blue galaxies in clusters is observed to increase with redshift. Here instead we consider the effect of environment on the fraction of blue galaxies. We define the blue galaxy fraction as the fraction of $R<20$ ($\sim$M$^{*}\!+\!2$) galaxies with rest-frame $B-V$ colours at least 0.2\,mag bluer than that of the C-M relation, as for the original studies of Butcher \& Oemler (\cite{bo}). We estimate the corresponding change in $B-R$ colour at the cluster redshift by firstly considering two model galaxies at $z=0$ (Bruzual \& Charlot \cite{bruzual}), one chosen to be a typical red-sequence galaxy (10\,Gyr old, $\tau=0.1$\,Gyr, $Z=0.02$), and the second reduced in age until its $B-V$ colour becomes 0.2\,magnitude bluer. After redshifting both galaxies to the cluster redshift, the difference in their observed $B-R$ colour is found to be 0.447\,mag. Hence we consider the blue galaxy fraction to be the fraction of $R<20$ galaxies having $B-R < 3.420 - 0.0815\times R$. Figure~\ref{blue} shows the resulting blue galaxy fractions in the high-, intermediate-, and low-density environments after averaging over 100 Monte-Carlo realisations of the cluster population. We find the blue galaxy fractions are low in both the high- ($0.062\pm0.042$) and intermediate-density ($0.088\pm0.042$) regions, but increases dramatically to $0.200\pm0.072$ in the low-density region. Kodama \& Bower (\cite{kodama}) examine the dependence of the blue galaxy fraction on radius for 7 clusters at $0.2<z<0.43$, and observe a general trend for $f_{B}$ to increase with radius, the effect becoming more prominent for the higher redshift clusters. For each cluster, the $f_{B}$ in the cluster core ($r\lsim0.5$\,Mpc), remains low at $\sim0.10$, whereas at cluster-centric radii of 1--2\,Mpc, the $f_{B}$ increases dramatically to 0.2--0.6.

\section{Galaxy colours}

\begin{figure}[t]
\centerline{{\resizebox{\hsize}{!}{\includegraphics{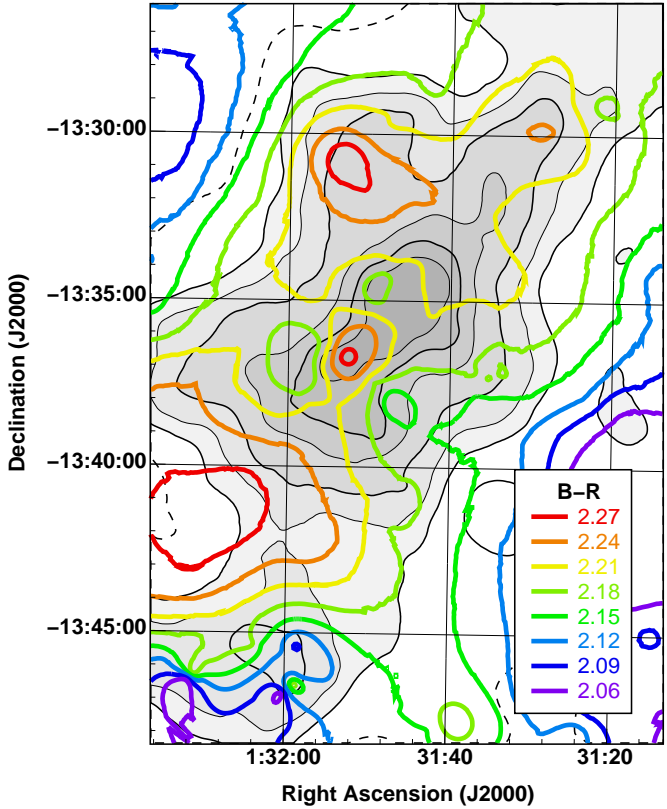}}}}
\caption{The mean galaxy colour of the cluster galaxy population as a function of spatial position (see text).}
\label{colour}
\end{figure}

To gain some further insight into the effect of cluster environment and also the particular dynamical state of A\,209, we plot in Figure~\ref{colour} the mean $B-R$ colour of $R<21$ cluster galaxies as a function of spatial position. This is determined as for the surface density map of Fig.~\ref{density} through the adaptive kernel approach (Pisani \cite{pisani93}; \cite{pisani96}), in which each galaxy is weighted according to the probability that it is a cluster member, $1-P(f)$, so that for a particular point, $\mathbf{x}$, the local mean galaxy colour is given by:
\begin{equation}
\overline{(B\!-\!R)}_{\mathbf{x}} = \frac{\sum_{i=1}^{N} (B\!-\!R)^{CM}_{i}.(1\!-\!P(f)_{i}).K(\mathbf{x}\!-\!\mathbf{x}_{i};r_{i})}{ \sum_{i=1}^{N} (1\!-\!P(f)_{i}).K(\mathbf{x}\!-\!\mathbf{x}_{i};r_{i})}
\end{equation}
where $\mathbf{x}_{i}$, $P(f)_{i}$ and $r_{i}$ are the position, probability of being a field galaxy, and kernel width respectively of galaxy $i$ out of $N$. And $(B-R)_{i}^{CM}$ is the $B-R$ colour for each galaxy after accounting for the luminosity-dependent effect of the slope of the CM-relation through
\begin{equation}
(B-R)_{i}^{CM}=(B-R)_{i}+0.0815 \times (R_{i}-19).
\end{equation}
The resultant mean galaxy colour map $[B\!-\!R](\mathbf{x})$ is shown by the coloured contours, with the red/orange contours indicating regions which have redder cluster galaxies on average, and blue/purple contours indicating regions which have more blue cluster galaxies. To see how the colour of these cluster galaxies are related to their environment, the contours are overlaid upon the same contour map of the surface number density of $R<23.0$ galaxies as shown in Fig.~\ref{density}.

What is most important is that only the cluster galaxy population is being examined, and so the fact that field galaxies are bluer on average than cluster galaxies should not result in a density-dependent colour gradient, as would be the case if all galaxies were considered. Instead, any surface density-dependent colour gradient should be purely the result of galaxies in the low-density environments on the cluster periphery having differing colours on average to those in high-density environments. Secondly, as it is the mean $B-R$ colour of $R<21$ galaxies that is being measured, it is the effect of the cluster environment on only the luminous cluster galaxies ($L\gsim0.1L^{*}$) that is being examined. 

A clear overall density-dependence on galaxy colour is apparent, with galaxies in the high-density regions having $\mu(B\!-\!R)\sim2$.21--2.30, while galaxies in the low-density regions have $\mu(B\!-\!R)\sim2$.14--2.18. This we take to be the manifestation of the same effect observed with the blue galaxy fraction, with higher blue galaxy fractions on the cluster peripheries, thus reducing the mean $B-R$ colour in those regions. %A second contributory effect will be the combination of the slope in the C-M relation, and the concentration of the most luminous red sequence galaxies towards the cluster core, which are redder than their fainter counterparts.

Substructure in the mean $B-R$ galaxy colours is apparent, and appears related to the dynamical state of the system, being aligned with the direction of elongation in the galaxy number surface density. The reddest mean $B-R$ galaxy colours are observed at the very centre of the cluster, coincident with the cD galaxy and a concentration of the brightest $R<19$ galaxies, as would be expected. There also appear regions of red galaxies on either side of the cluster core aligned with the overall elongation of the cluster. As the overall galaxy surface density is greater in the NW direction from the cluster core, we would suggest that this is where the merging clump is found, some 4--5\,arcmin from the centre of A\,209, corresponding to $\sim1$\,Mpc at $z=0.209$.

In contrast, perpendicular to the axis of elongation appear regions close to the cluster centre containing bluer than average galaxies. The concentration to the east appears centred on a bright face-on spiral (\mbox{$R=17.71$}, \mbox{$B-R=1.47$}), which has been spectroscopically confirmed as a cluster member (Mercurio et al. \cite{paper1}).
  
\section{DISCUSSION}

We have examined the effect of cluster environment, as measured in terms of the local surface density of \mbox{$R<23.0$} galaxies, on the global properties of the cluster galaxies, through their luminosity functions, colour-magnitude relations, and average colours. For this study we have considered three cluster environments, a high-density region sampling the cluster core, and intermediate- and low-density regions which sample the cluster periphery. 

\subsection{The Galaxy Luminosity Function}

The LFs for galaxies within the virialised region are found to be well described by single Schechter functions to M$_{R}$, M$_{B}\sim-16$, although there is an indication of a dip at $R=2$0--20.5 (M$_{R}=-20$). Dips in the LF at such absolute magnitudes appear common for low-redshift clusters (see Mercurio \etal \cite{paper2}), and appear stronger for richer, early-type dominated clusters (Yagi \etal \cite{yagi}). %By considering the bright (E/S0/Sabc) and dwarf cluster galaxy populations separately, through type-specific luminosity functions (TSLFs), it becomes apparent that the dip is the byproduct of the two populations having significantly differing TSLFs. The bright galaxy LF can be fitted by a Schechter function with $M_{R}^{*}\sim-21.5$ and $\alpha=-0.$5--1.0, whereas the dwarf galaxy LF is much fainter (\mbox{$M_{R}^{*}\sim-17.5$} and has a much steeper faint-end slope, \mbox{$\alpha\sim-1.40$} (e.g. Binggeli \etal \cite{binggeli}; Yagi \etal \cite{yagi}; Driver \etal \cite{driver}). The dip in the total LF then occurs at a magnitude where the bright galaxy population is beginning to tail off, and brighter than that where the dwarf galaxies begin to dominate.

The faint-end slope, $\alpha$, shows a strong dependence on environment, becoming steeper at $>3\sigma$ significance level from high- to low-density environments. We explain this trend as the combination of two related effects:  a manifestation of the morphology-density relation whereby the fraction of early-type galaxies which have shallow faint-end slopes increases with density, at the expense of late-type galaxies which have steep faint-end slopes; and a luminosity-segregation due to dwarf galaxies being cannibalised and disrupted by the cD galaxy and the ICM in the cluster core \cite{lopezcruz}.

To separate the two effects, we consider the TSLF of galaxies belonging to the cluster red sequence, as these are predominately early-type galaxies, and so any trends with environment should be independent of the morphology-density relation. Some luminosity-segregation is observed, with a reduction in the fraction of dwarf galaxies in the high-density regions, as manifested by their shallower TSLF faint-end slopes. This effect is smaller than that for the overall LF, being significant only at the $1.7\sigma$ level, indicating that both luminosity-segregation and the morphology-density relation drive the observed trends in the overall LF. Luminosity-segregation is predicted by simulations for early-type galaxies in clusters, with bright early-type galaxies much more concentrated than their faint counterparts which follow the distribution of the dark matter mass profile (Springel \etal \cite{springel}). 

In a study of 45 low-redshift \mbox{$(0.04<z<0.18)$ }clusters, L\'{o}pez-Cruz \etal (\cite{lopezcruz}) find that for the seven that are rich, dynamically-evolved clusters, characterised by the presence of a cD galaxy, their composite LFs are all well described by single Schechter functions with shallow faint-ends \mbox{($\alpha\approx$-1.0)}. In contrast other clusters, often poorer, but in particular those not containing a cD galaxy, require two Schechter functions to fit their LFs, including a steep faint-end slope to model the dwarf population (see also Parolin, Molinari \& Chincarini \cite{parolin}). We thus indicate that the shallow faint-end slope observed in the high-density region of A\,209 is related to the presence of the central dominant galaxy. cD galaxies are regarded as physically different to elliptical galaxies, with a different formation history, and are the product of dynamic processes which take place during the formation of their host cluster. They are thought to be built up through galactic cannibalism, or the accumulation of tidal debris. L\'{o}pez-Cruz \etal (\cite{lopezcruz}) propose that the flatness of the faint-end slope in clusters containing cD galaxies results from the disruption of a large fraction of dwarf galaxies during the early stages of cluster evolution, the stars and gas of which are cannibalised by the cD galaxy, and the remainder redistributed into the intracluster medium.
 
\subsection{Environmental Effects on the Red Sequence}

As well as having an affect on the TSLF of red sequence galaxies, the cluster environment could have an affect on the colour or slope of the red sequence itself, through the mean ages or metallicities of the galaxies. To examine this possibility the red sequence was fitted for the high- and intermediate-density regions independently. The red sequence was found to be \mbox{$0.022\pm0.014$\,mag} redder in the high-density region than for the intermediate-density region by fixing the slope. In contrast no correlation between the slope of the red sequence and environment was observed. A similar effect is observed for bulge-dominated galaxies taken from the Sloan Digital Sky Survey (SDSS) (Hogg et al. 2003), in which the modal $^{0.1}[g-r]$ residual colour to the best-fitting C-M relation is \mbox{0.01--0.02\,mag} redder in the highest-density environments (corresponding to cluster cores) than in their low-density counterparts. In a study of 11 X-ray luminous clusters at $0.07<z<0.16$ Pimbblet \etal (\cite{pimbblet}) examine the red sequence as a function of cluster-centric radius and local galaxy density, and observe that the relation becomes progressively bluer as cluster-centric radius is increased out to 3\,Mpc, and as the local surface density is decreased at a rate of $d(B-R)/d\log_{10}(\Sigma)=-0.08\pm0.01$.
% However they suggest that as the dispersion of the relation also increases with radius, the trends are due to an increasing fraction of blue galaxies among the red sequence galaxies, rather than a change in the colours of the whole population. 
In a study of the cluster A\,2390 at $z=0.23$ Abraham \etal (\cite{abraham}) also observe the normalised colour of red sequence galaxies, $(g-r)_{r=19}$, to become increasingly blue with increasing cluster-centric radius, $r_{p}$, as \mbox{$(g-r)_{r=19}=1.05-0.079\log r_{p}$}.

These results indicate that the environment can affect the colour of a red sequence galaxy, so that red sequence galaxies are older and/or have higher metallicities in denser environments. To quantify this effect we consider a model red sequence galaxy as a 7.5\,Gyr old  (at \mbox{$z=0.209$}, corresponding to 10\,Gyr at the present epoch), \mbox{$\tau=0.01$\,Gyr}, solar-metallicity stellar population (Bruzual \& Charlot \cite{bruzual}). By varying its age and metallicity independently, we find that to reproduce the observed reddening in the C-M relation, galaxies in the high-density region must be on average 500\,Myr older or 20\% more metal-rich than their intermediate-density counterparts. By studying the spectra of 22\,000 luminous, red, bulge-dominated galaxies from the SDSS, Eisenstein \etal (\cite{eisenstein}) indicate that red sequence galaxies in high-density regions are marginally older and more metal-rich than their counterparts in low-density environments, with both effects coming in at the same level. 

This result is understandable in terms of cosmological models of structure formation, in which galaxies form earliest in the highest-density regions corresponding to the cores of rich clusters. Not only do the galaxies form earliest here, but the bulk of their star-formation is complete by $z\sim1$, by which point the cluster core is filled by shock-heated virialised gas which does not easily cool or collapse, suppressing the further formation of stars and galaxies (Blanton \etal \cite{blanton99}; \cite{blanton00}). Diaferio \etal (\cite{diaferio}) shows that as mixing of the galaxy population is incomplete during cluster assembly, the positions of galaxies within the cluster are correlated with the epoch at which they were accreted. Hence galaxies in the cluster periphery are accreted later, and so have their star-formation suppressed later, resulting in younger mean stellar populations. It should be considered however that this is a small-scale affect, and that this result in fact confirms that the red sequence galaxy population is remarkably homogeneous across all environments.

\subsection{Galaxy Colours}

As well as considering the effect of the cluster environment on galaxy morphologies and LFs, its effect on star-formation has been examined by numerous authors (e.g. Balogh \etal \cite{balogh}; Ellingson \etal \cite{ellingson}; Lewis \etal \cite{lewis}). They find that star-formation is consistently suppressed relative to field levels for cluster galaxies as far as twice the virial radius from the cluster centre, and that for the majority of galaxies in the cluster core star-formation is virtually zero. For photometric studies it is possible to qualitatively measure the effect of the cluster environment on star-formation through measurement of the blue galaxy fraction --- the fraction of luminous \mbox{($M_{R}^{*}<-20$)} galaxies whose colours indicate they are undergoing star-formation typical of late-type galaxies. We find that the blue galaxy fraction decreases monotonically with density, in agreement with other studies (e.g. Abraham \etal \cite{abraham}; Kodama \& Bower \cite{kodama}).

The observed trends of steepening of the faint-end slope, faintening of the characteristic luminosity, and increasing blue galaxy fraction, from high- to low-density environments, are all manifestations of the well known morphology-density relation (Dressler \cite{dressler80}; Dressler \etal \cite{dressler97}), where the fraction of early-type galaxies decreases smoothly and monotonically from the cluster core to the periphery, while the fraction of late-type galaxies increases in the same manner. The observed trends in the composite LF simply reflect this morphology-density relation: the galaxy population in the cluster core is dominated by early-type galaxies and so the composite LF resembles that of this type of galaxy, with a shallow faint-end slope and a bright characteristic luminosity; whereas in lower density regions the fraction of late-type galaxies increases, and so the composite LF increasingly resembles that of the late-type TSLF, with a steep faint-end slope and a fainter characteristic magnitude (Binggeli \etal \cite{binggeli}).  

Finally, we examined the effect of the cluster environment on galaxies through measuring the mean colour of luminous (\mbox{$R<21$}) cluster galaxies as a function of their spatial position, as shown in Fig.~\ref{colour}. This shows more clearly than any other result, the complex effects of the cluster environment and dynamics on their constituent galaxies. A\,209 appears a dynamically young cluster, with a significant elongation in the SE-NW direction, the result of a recent merger with smaller clumps. To measure the effect of the cluster dynamics on the galaxy population, their properties should be measured against the parameter most easily related back to the cluster dynamics, their spatial position. As the cluster is significantly elongated, it should be possible to distinguish between whether the local density or cluster-centric distance is more important in defining the properties of galaxies, and indeed it is clear that the mean galaxy colour correlates most with the local density rather than the distance of the galaxy from the cluster core. It should be considered through that as the system appears elongated due to being the merger of two or more clumps, the notion of a cluster-centric radius loses much of its validity, as where does the centre of the system lie during a cluster merger? The location of the main cluster and the secondary merging clump appear confirmed by Fig.~\ref{colour}, with the reddest galaxies concentrated around the cD galaxy (main cluster) and a more diffuse region 5\,arcmin to the north coincident with the structure predicted from weak lensing analysis (Dahle \etal \cite{dahle}).
 The effect of the preferential SE-NW direction for A\,209 is apparent in the presence of bright blue galaxies near the cD galaxy perpendicular to the axis and hence unaffected by the cluster merger, and an extension of red galaxies to the SE which may indicate the infall of galaxies into the cluster along a filament. This preferential SE-NW direction appears related to the large-scale structure in which A209 is embedded, with two rich (Abell class R=3) clusters A\,222 at \mbox{$z=0.211$} and A\,223 at \mbox{$z=0.2070$} are located \mbox{$1.5^{\circ}$} (15\,Mpc) to the NW along this preferential axis.

Cluster dynamics and large-scale structure clearly have a strong influence on galaxy evolution, and it would be interesting to search for direct evidence of their effect on the star-formation histories of galaxies through spectroscopic observations of galaxies in the secondary clump, in the form of post-starburst signatures.

\begin{acknowledgements}
CPH acknowledges the financial support provided through the European Community's Human Potential Program under contract HPRN-CT-2002-00316, SISCO.
\end{acknowledgements}

\end{document}